\documentclass{PoS}

\usepackage{graphicx}

\title{Dirac $b$ quark on the lattice}

\def\beq{\begin{equation}}
\def\eeq{\end{equation}}
\def\bea{\begin{eqnarray}}
\def\eea{\end{eqnarray}}
\def\bean{\begin{eqnarray*}}
\def\eean{\end{eqnarray*}}
\def\Id{\mbox{1\hspace{-1.0mm}I}}
\def\tr{\mathrm{tr}}
\def\b{\mathbf{b}}
\def\c{\mathbf{c}}
\def\s{\mathbf{s}}
\def\u{\mathbf{u}}
\def\d{\mathbf{d}}

\def\bbar{\bar\mathbf{b}}
\def\cbar{\bar\mathbf{c}}

\def\Qbar{\bar\mathbf{Q}}

\ShortTitle{Dirac $b$ quark on the lattice}

\author{TWQCD Collaboration:
	Ting-Wai Chiu$^a$,
	\speaker{Tung-Han Hsieh}$^b$,
	Chao-Hsi Huang$^c$,
	and Kenji Ogawa$^d$\\
$^a$ Department of Physics, Center for Theoretical Sciences,  
     and National Center for Theoretical \\
     \hspace{2mm} Sciences,
     National Taiwan University, Taipei 10617, Taiwan.\\
     \hspace{2mm}Email: \email{twchiu@phys.ntu.edu.tw}\\
$^b$ Research Center for Applied Sciences,
     Academia Sinica, Taipei 115, Taiwan.\\
     \hspace{2mm}Email: \email{thhsieh@twcp1.phys.ntu.edu.tw}\\
$^c$ Department of Physics,
     National Taiwan University, Taipei 10617, Taiwan.\\
     \hspace{2mm}Email: \email{chao@twcp1.phys.ntu.edu.tw}\\
$^d$ Department of Physics,
     National Taiwan University, Taipei 10617, Taiwan.\\
     \hspace{2mm}Email: \email{ogawak@phys.ntu.edu.tw}%
}

\abstract{%
We perform the first study of treating $ \b $, $ \c $, and $ \s $ quarks
as Dirac fermions in lattice QCD with exact chiral symmetry.
On a $ 32^3 \times 60 $ lattice with $ a^{-1} \simeq 7.68 $ GeV,
we compute point-to-point quark propagators,
and measure the time-correlation functions for mesons
with quark contents $ \b\bbar $, $ \c\bbar $, $\s\bbar $, and $ \c\cbar $.
The lowest-lying meson mass spectra, the pseudoscalar decay constants,
and the $ \b $ and $ \c $ quark masses are determined.%
}

\FullConference{The XXV International Symposium on Lattice Field Theory\\
		 July 30-4 August 2007\\
		 Regensburg, Germany}

\begin{document}


\section{Introduction}

Spectroscopy with heavy quark is one of the major topics 
in high energy physics. 
Since the $ \b $ quark is very heavy ($\sim 4.6 $ GeV), 
most theoretical approaches regard the $ \b $ quark as a static, 
non-relativistic particle, or treat it with heavy quark effective
theory or the relativistic heavy quark formalism. In other words, 
the heavy $ \b $ quark is not treated as the Dirac fermion. 
This introduces complicated normalization
procedures, in which the systematic errors are difficult to control.

In QCD, all quarks are excitations of Dirac fermion fields, thus it is
vital to preserve this important feature in any theoretical
study. However, so far, it remains a challenge for a single 
lattice to accommodate such a wide
range of quark masses from $m_{u/d}$ quark to $ m_b $. 
If we restrcit our study to the hadrons containing only
$\s$, $\c$, and $\b$ valence quarks (ranging from 140 MeV to 4.6 GeV) 
satisfying the constraints $m_q a < 1$ and $M_h L > 4$, then it is possible 
to accommodate these hadrons on a $32^3 \times 60$ lattice, 
with the inverse of lattice spacing $a^{-1} \simeq 7.68$ GeV.

In this proceeding, we report the first study of 
heavy meson spectra \cite{BeautyMesons}, 
by treating all heavy and light quarks as Dirac
fermions, in lattice QCD with exact chiral symmetry. 
Namely, we restrict ourselves to mesons with quark contents 
$\b\bbar$, $\c\bbar$, $\s\bbar$, and $\c\cbar$. 
Our results of masses and decay constants of
the pseudoscalar mesons $B_s$ and $B_c$, and the masses of the vector
mesons $B_s^*$ and $B_c^*$, have been presented in \cite{Chiu:2007bc}.

\begin{figure}[htb]
\begin{center}
\begin{tabular}{@{}cc@{}}
\includegraphics*[height=8cm,width=7cm]{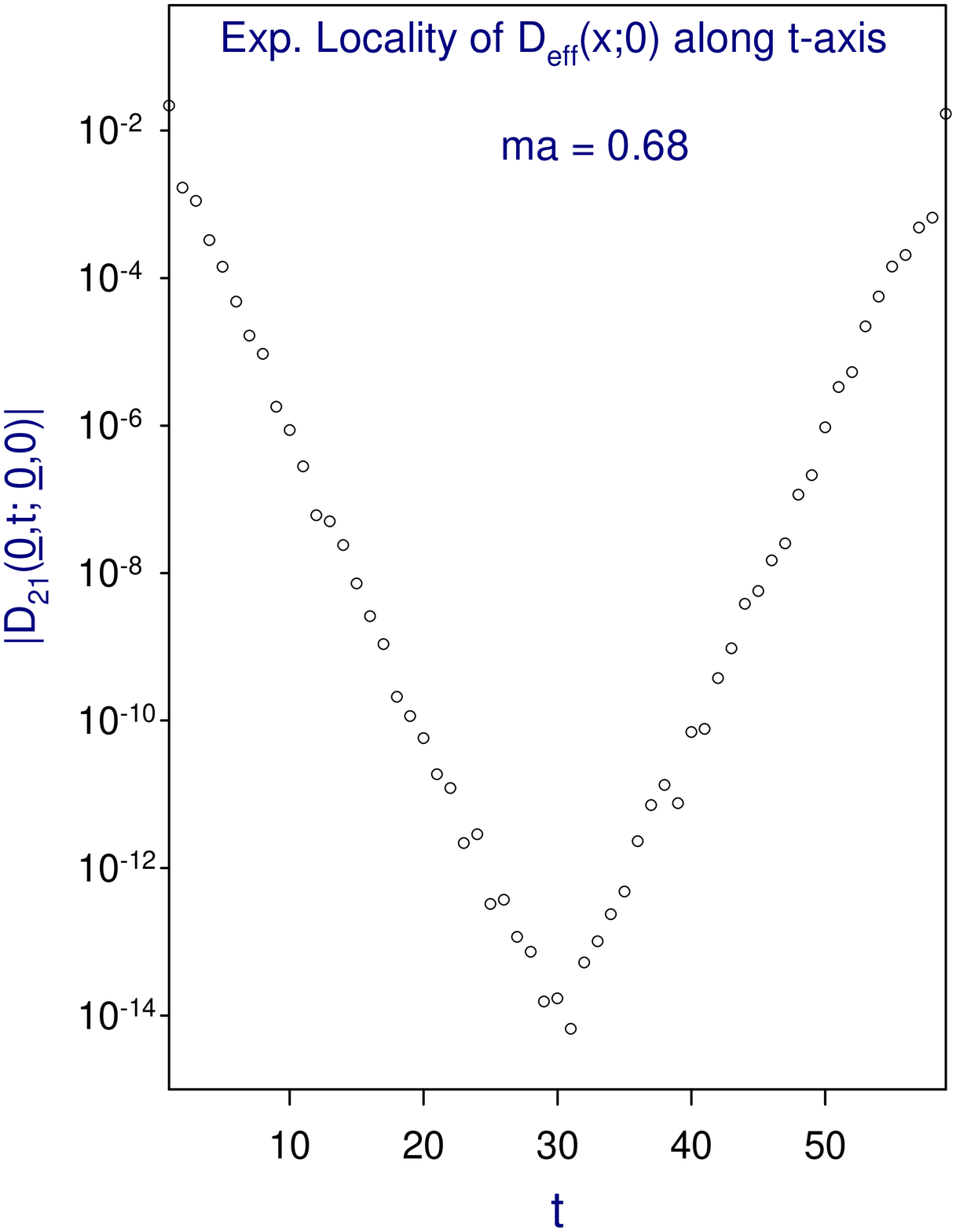}
&
\includegraphics*[height=8cm,width=7cm]{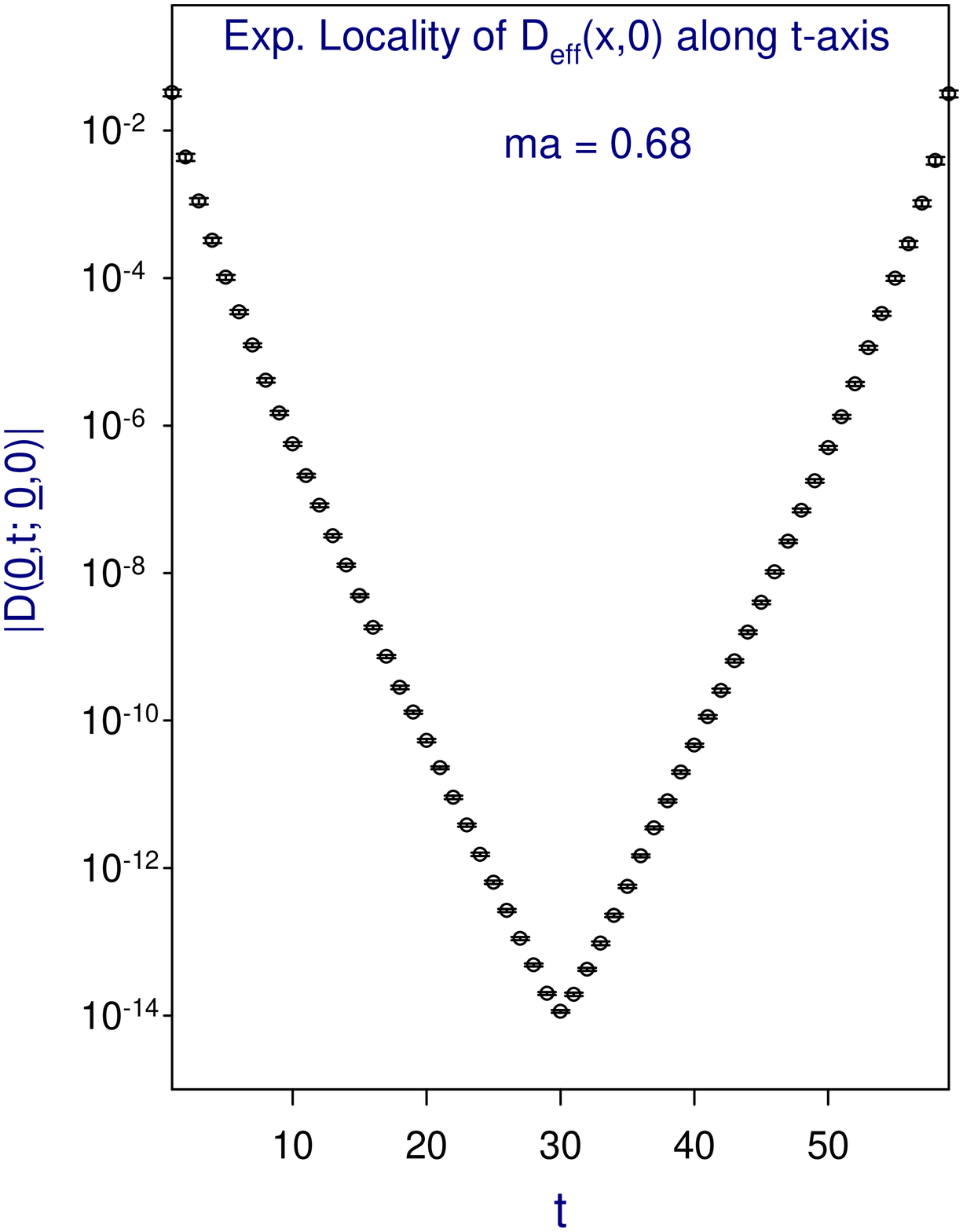}
\\ (a) & (b)
\end{tabular}
\caption{
(a) 
The magnitude of the an element of $ D_{\mbox{eff}}(\vec{0},t;\vec{0},0) $ 
(with color index =2, and Dirac index = 1) 
from the origin to any point along $t$-axis, for one configuration, 
at $m_q a = m_b a = 0.68$.
(b) Similar to (a), but averaging over all elements with different 
color and Dirac indices.
}
\label{fig:locality}
\end{center}
\end{figure}

We adopt the optimal domain-wall fermion (ODWF) proposed by Chiu
\cite{Chiu:2002ir, Chiu:2003ir} as our fermion scheme, which possesses 
optimal chiral symmetry for any finite $ N_s $ (the number of sites in the 
5-th dimension), and also preserves the salient features of 
the Dirac fermion in the continuum. 
With 100 gauge configurations generated with the single plaquette action at
$\beta=7.2$, we compute the point-to-point quark propagators for 33 quark
masses in the range $0.01 \le m_q a \le 0.85$. We set the fifth dimension
of ODWF, $N_s$, to 128, and the stopping criteria of the conjugate gradient
(CG) iterations to $10^{-11}$, such that the chiral symmetry breaking
$\sigma$ and the residual of the quark propagators $\epsilon$ satisfy 
\bean
\sigma = |Y^{\dagger} S^2 Y| / | Y^{\dagger} Y | < 10^{-14}, \hspace{4mm}
\epsilon = \| (D_c+m_q) Y -\Id \| < 2\times 10^{-11}
\eean
where $S$ is the optimal rational approximation of the sign
function of $\gamma_5 D_w$, $D_w$ is the Wilson Dirac operator 
plus a negative parameter $ -m_0 $. Here
$(D_c+m_q)^{-1}$ is the valence quark propagator, and $Y$ is the
solution vector of solving the quark propagators via the CG algorithm.
Note that our massive Dirac operator is exponentially local for all the quark
masses. In Fig.\ \ref{fig:locality}, we plot the magnitude
of the 4-D effective Dirac operator of ODWF, $ D_{\mbox{eff}}(x;0) $ 
(i.e., the overlap-Dirac operator with Zolotarev approximation 
of the sign function) \cite{Chiu:2003ir} along the t-axis, 
for one gauge configuration. Evidently, it is exponentially-local. 
Furthermore, we have checked that for any element of 
$ D_{\mbox{eff}}^{a\alpha;b\beta}(x;0) $ 
(where $ a,b=1,2,3 $ and $ \alpha,\beta=1,2,3,4 $), its magnitude 
is also exponentially-local along any direction, 
for all gauge configurations and all 33 quark masses.

Next, we measure the time-correlation function
\beq
\label{eq:TCF_meson}
C_{\Gamma}(t) = \left< \sum_{\vec x}\tr\{ \Gamma(D_c+m_Q)_{x,0}^{-1}
					  \Gamma(D_c+m_q)_{0,x}^{-1} \}\right>
\eeq
for scalar ($S$), pseudoscalar ($P$), vector ($V$), axial-vector ($A$), and
tensor ($T$) mesons, with Dirac matrix $\Gamma = \{ \Id, \gamma_5, \gamma_i,
\gamma_5\gamma_i, \gamma_5\gamma_4\gamma_i \}$ respectively, for both
symmetric ($m_Q = m_q$) and asymmetric (fixed $m_Q$ with various $m_q$)
quark masses. Finally we extract their masses and pseudoscalar decay constants,
compare our results with the experimental data, and also make some theoretical
predictions.


\section{Determination of $a^{-1}$, $m_c$, $m_s$, and $m_b$}

In Ref.\ \cite{Chiu:2005ue}, we determine the inverse lattice
spacing from the pion decay constant, with the experimental input
$f_{\pi} = 131$ MeV. However, in this work, our smallest quark mass
turns out to be too heavy (about $m_s/2$), thus the chiral extrapolation
to $m_q \simeq 0$ does not seem to be feasible. Instead, we use the mass
and decay constant of the pseudoscalar meson $\eta_c(2980)$ to determine
the bare mass of charm quark $m_c$ and $a^{-1}$ simultaneously, which can
be seen in the following.

We measure the time-corelation function Eq.(\ref{eq:TCF_meson}) of the
pseudoscalar meson $C_P(t)$ ($\Gamma=\gamma_5$) for symmetric quark mass
$m_Q = m_q$, and fit it to the usual formula $G(t)$, to extract the meson
mass $m_P a$ and the decay constant $f_P a$:
\beq
\label{eq:fit_meson}
G(t) = \frac{z^2}{2 m_P a}[\, e^{-m_P a t}+e^{-m_P a (T-t)}\,],
\hspace{10mm}
f_P a = 2 m_q a \frac{z}{m_P^2 a^2}
\eeq
Thus, we obtain the ratio of $m_P / f_P$ for each $m_q a$ we have computed,
which can be compared with the ratio of mass and decay constant of $\eta_c$.

\begin{table}[th]
\begin{center}
\begin{tabular}{c|cccc}
\hline
 &
\raisebox{-0.5mm}{lattice} &
\raisebox{-0.5mm}{$\overline{\mathrm{MS}}$} &
\raisebox{-0.5mm}{$\overline{\mathrm{MS}}$} &
\raisebox{-0.5mm}{PDG}
\\
\raisebox{0.5mm}{$m_q$} &
\raisebox{0.5mm}{bare mass} &
\raisebox{0.5mm}{(at $\mu=2$ GeV)} &
\raisebox{0.5mm}{(at $\mu=m_q$)} &
\raisebox{0.5mm}{average} \\
\hline
$m_b$ & 5.22(4) GeV & 4.81(5) GeV & 4.65(5) GeV & 4.20(7) GeV \\
$m_c$ & 1.23(4) GeV & 1.13(4) GeV & 1.16(4) GeV & 1.25(9) GeV \\
\hline
\end{tabular}
\end{center}
\caption{\label{tab:mb_mc_mass}
The bare mass ($m_q a^{-1}$) of $\b$ and $\c$ quarks, and their
masses in $\overline{\mathrm{MS}}$ scheme
at scales $\mu = 2$ GeV and $\mu = m_q$, respectively, in comparison 
with the PDG average in the last column.%
}
\end{table}

Although only the $m_{\eta_c}$ is measured in the high energy experiments,
we can obtain the theoretical value of $m_{\eta_c}/f_{\eta_c} = 6.8(2)$ from
our previous study of pseudoscalar mesons on the $20^3 \times 40$ lattice at
$\beta=6.1$ \cite{Chiu:2005ue}. We find that the closest value of
$m_P / f_P = 6.8(1)$ occurs at $m_q a=0.16$, thus we fix $m_c a=0.16$.
Further, using the experimental input $m_{\eta_c} = 2980$ MeV, we determine
the inverse of the lattice spacing $a^{-1} = 7680(59)$ MeV. To check the
goodness of our determination, we measure the time-correlation function of
the vector meson $\cbar\gamma_i\c$, and extract its mass equal to 3091(11) MeV,
which is in good agreement with $J/\Psi (3097)$.

Next, we determine the bare masses of strange and bottom quark, $m_s a$ and
$m_b a$, by extracting the mass of vector meson from the time-correlation
function $C_V(t)$ for various $m_q a$ we have computed. At $m_q a = 0.02$,
we obtain $m_V = 1027(38)$ MeV, in good agreement with $\phi(1020)$; while
at $m_q a = 0.68$, we obtain $m_V = 9453(3)$ MeV, in good agreement with
$\Gamma(9460)$. Thus we determine $m_s a = 0.02$ and $m_b a=0.68$,
respectively.

Note that although the spatial size of our lattice ($L\simeq 0.8$ fm)
seems to be small, the finite size effects should be well under control since
our pseudoscalar meson masses satisfies $m_P L > 4$ even for our smallest
bare quark mass $m_q a = 0.01$.

In order to compare our results of $m_b$ and $m_c$ with the high energy
phenomenology, we have to obtain the lattice renormalization constants
$Z_m = Z_s^{-1}$ and transcribe them to the usual $\overline{\mathrm{MS}}$
scheme, where $Z_s$ is the renormalization constant for $\psi\bar\psi$. In
general, the renormalization constants should be determined nonperturbatively.
However, in this work, the lattice spacing is rather small 
($a \simeq 0.026$ fm),
we suspect that the one-loop perturbation formula \cite{Alexandrou:2000kj}
\beq
\label{eq:normalization:Z_s}
Z_s(\mu) = 1 + \frac{g^2}{4\pi^2}[ \ln(a^2\mu^2) + 0.17154 ],
\hspace{10mm} (m_0=1.30)
\eeq
already provides a very good approximation for $Z_s$. At $\beta=7.2$,
$a^{-1} = 7.680(59)$ GeV, our results for $m_b^{\overline{\mathrm{MS}}}$ and
$m_c^{\overline{\mathrm{MS}}}$ at scales $\mu = 2$ GeV and their masses
($\mu = m_q$) are listed in Table \ref{tab:mb_mc_mass}. Comparing with the PDG
\cite{Yao:2006px} average, our result of $m_b^{\overline{\mathrm{MS}}}(m_b)$
turns out to be a little higher, but $m_c^{\overline{\mathrm{MS}}}(m_c)$ is
in good agreement with the PDG average.


\section{Charmonium $\c\cbar$ and Bottomonium $\b\bbar$}

\begin{table}[th]
\begin{center}
\begin{tabular}{@{}l@{\ \ \ }llllll@{}}
\hline
$ \Gamma $ & $ J^{PC} $ & n$^{2S+1} L_J $ 
           & $ [t_{min},t_{max}] $ & $\chi^2$/dof & Mass(MeV) & PDG \\
\hline
$ \Id $ & $ 0^{++} $ & $ 1^3 P_0 $ 
                     & [19,42] & 0.32 
                     & 3413(14)(9) 
                     & $ \chi_{c0}(3415) $ \\
$ \gamma_5 $ & $ 0^{-+} $ & $ 1^1 S_0 $ 
                          & [22,38] & 1.02 
                          & 2980(10)(12) 
                          & $ \eta_c(2980) $ \\ 
$ \gamma_i $ & $ 1^{--} $ & $ 1^3 S_1 $ 
                          & [19,38] & 1.18 
                          & 3091(11)(14) 
                          & $ J/\psi(3097) $ \\
$ \gamma_5\gamma_i $ & $ 1^{++} $ & $ 1^3 P_1 $ 
                                  & [18,42] & 0.51 
                                  & 3516(13)(8) 
                                  & $ \chi_{c1}(3510) $ \\
$ \gamma_5\gamma_4\gamma_i $ & $ 1^{+-} $ & $ 1^1 P_1 $ 
                             & [19,43] & 0.41 
                             & 3526(13)(9) 
                             & $h_c(3524)$ \\
\hline
\end{tabular}
\end{center}
\caption{\label{tab:ccbar}
The mass spectra of the lowest-lying ($n=1$) charmonium $\cbar\Gamma\c$ states
obtained in this work, in comparison with the PDG values in the last column.}
\end{table}

\begin{table}[th]
\begin{center}
\begin{tabular}{@{}lllllll@{}}
\hline
$ \Gamma $ & $ J^{PC} $ & n$^{2S+1} L_J $ 
           & $ [t_{min},t_{max}] $ & $\chi^2$/dof & Mass(MeV) & PDG \\
\hline
%
%
$ \Id $ & $ 0^{++} $ & $ 1^3 P_0 $ 
                     & [21,39] & 0.21 
                     & 9863(15)(8) 
                     & $ \chi_{b0}(9859) $ \\
%
$\gamma_5$ & $ 0^{-+} $ & $ 1^1 S_0 $ 
                        & [27,35] & 0.72 
                        & 9383(4)(2) 
                        & $ \eta_b(9300) $ ? \\
%
%
%
%
%
$\gamma_i$ & $ 1^{--} $ & $ 1^3 S_1 $ 
                        & [20,39] & 1.19 
                        & 9453(3)(2) 
                        & $ \Upsilon(9460) $ \\
%
%
%
%
$\gamma_5\gamma_i$ & $ 1^{++}$ & $1^3 P_1$ 
                               & [22,38] & 0.13 
                               & 9896(20)(8) 
                               & $ \chi_{b1}(9893) $ \\
%
%
$\gamma_5\gamma_4\gamma_i$ & $ 1^{+-} $ & $ 1^1 P_1 $ 
                           & [22,38] & 0.10 
                           & 9916(22)(8) 
                           &  \\
\hline 
\end{tabular}
\end{center}
\caption{\label{tab:bbbar}
The mass spectra of the lowest-lying ($n=1$) bottomonium $\bbar\Gamma\b$ states
obtained in this work. The last column is the experimental state we have
identified, and its PDG mass value.}
\end{table}

First of all, we check to what extent we can reproduce the charmonium mass
spectra which have been measured precisely by high energy experiments.
Our results of the mass spectra of the lowest-lying states of charmonium 
are summarized in Table \ref{tab:ccbar}. The first column is
the gamma matrix we use to compute the meson time-correlation function
(Eq.(\ref{eq:TCF_meson}). The second and the third columns are the $J^{PC}$
of the state and the conventional spectrascopic notation. The fourth column 
is the time range for fitting the time-correlation function to 
the formula in Eq.(\ref{eq:fit_meson}). The $\chi^2/\mathrm{dof}$ is listed
in the fifth column, and the extracted meson mass is listed in the sixth
column, where the first error is the statistical error, and the second
error is the estimated systematic error based on all fittings satisfying
$\chi^2/\mathrm{dof} < 1.3$ and $|t_{\mathrm{max}}-t_{\mathrm{min}}|\ge 6$,
with $t_{\mathrm{min}}\ge 10$ and $t_{\mathrm{max}} \le 50$. The last
column is the corresponding state in high energy experiments, with the
PDG mass value. Evidently, our results are in good agreement with the
PDG values. Note that our result of the hyperfine splitting
$(1^3 S_1 - 1^1 S_0)$ is 111(14)(18) MeV, comparing with the PDG value
118 MeV.

For the pseudoscalar $\eta_c$, we also obtain its decay constant
$f_{\eta_c}$ together with its mass through Eq.(\ref{eq:fit_meson}).
Our result is $f_{\eta_c} = 438\pm 5\pm 6$ MeV, where the first error
is the statistical error, and the second error is the estimated
systematic error, as described previously.

Next, we turn to the bottomonium ($\b\bbar$) states. Our results of the
mass spectra of its lowest-lying states are summarized in
Table \ref{tab:bbbar}. For the pseudoscalar $\eta_b$, it was first reported
by ALEPH Collaboration \cite{Heister:2002if}, but it has not been confirmed
by other HEP experimental groups. Our theoretical values turns to be
a little deviate from their experimental result. In addition, we also
determined its decay constant $f_{\eta_b} = 801 \pm 7 \pm 5$ MeV.

Finally, we note that the tensor meson $h_b$ has not been observed in
high energy experiments, thus our result of its mass serves as the first
prediction from lattice QCD with exact chiral symmetry.


\section{Mesons with quark contents $\s\bbar$ and $\c\bbar$}

\begin{table}[th]
\begin{center}
\begin{tabular}{@{}lllllll@{}}
\hline
$ \Gamma $ & $ J^{P} $ & n$^{2S+1} L_J $ 
           & $ [t_{min},t_{max}] $ & $\chi^2$/dof & Mass(MeV) & PDG \\
\hline
%
%
$ \Id $ & $ 0^{+} $ & $ 1^3 P_0 $ 
                    & [20,40]  & 0.38 
                    & 5852(15)(12) 
                    & $ B_{sJ}^* (5850) $  \\
%
$\gamma_5$ & $0^{-}$ & $ 1^1 S_0 $  
           & [26,32] & 0.56 
           & 5385(27)(17) 
           & $ B_s(5368) $ \\ 
%
%
%
%
%
$\gamma_i$ & $ 1^{-} $ & $ 1^3 S_1 $ 
           & [25,33] & 0.50 
           & 5424(28)(19) 
           & $ B_s^*(5412)$ \\ 
%
%
%
%
$\gamma_5\gamma_i$ & $ 1^{+} $ & $1^3 P_1$ 
                               & [18,42] & 0.68 
                               & 5884(16)(13) 
                               &  \\
%
%
$\gamma_5\gamma_4\gamma_i$ & $ 1^{+} $ & $ 1^1 P_1 $ 
                           & [18,38] & 0.62 
                           & 5897(16)(12) 
                           &    \\
\hline 
\end{tabular}
\end{center}
\caption{\label{tab:sbbar}
The mass spectra of the lowest-lying $\bbar\Gamma\s$ meson states obtained
in this work. The last column is the experimental state we have identified,
and its PDG mass value.}
\end{table}

\begin{table}[th]
\begin{center}
\begin{tabular}{@{}lllllll@{}}
\hline
$ \Gamma $ & $ J^{P} $ & n$^{2S+1} L_J $ 
           & $ [t_{min},t_{max}] $ & $\chi^2$/dof & Mass(MeV) & PDG \\
\hline
%
%
$ \Id $ & $ 0^{+} $ & $ 1^3 P_0 $ 
                    & [19,41] & 0.26 
                    & 6732(13)(9) 
                    &  \\
%
$\gamma_5$ & $ 0^{-} $ & $ 1^1 S_0 $ 
                       & [19,38] & 1.14 
                       & 6278(6)(4) 
                       & $ B_c(6286) $ \\
%
%
%
%
%
$\gamma_i$ & $ 1^{-} $ & $ 1^3 S_1 $ 
                       & [19,38] & 1.31 
                       & 6315(6)(5) 
                       & \\ 
%
%
%
%
$\gamma_5\gamma_i$ &$ 1^{+}$ & $ 1^3 P_1$ 
                             & [19,41] & 0.27  
                             & 6778(12)(7) 
                             &   \\
%
%
$\gamma_5\gamma_4\gamma_i$ & $ 1^{+} $ & $ 1^1 P_1 $ 
                           & [18,42] & 0.45 
                           & 6796(10)(7) 
                           &    \\
\hline 
\end{tabular}
\end{center}
\caption{\label{tab:cbbar}
The mass spectra of the lowest-lying $\bbar\Gamma\c$ meson states obtained
in this work. The last column is the experimental state we have identified,
and its PDG mass value.}
\end{table}


The unitarity of CKM matrix is one of the crucial tests for the Standard
Model. In order to extract the CKM matrix elements, the decay constants
of heavy-light pseudoscalar mesons (e.g., $f_B$, $f_{B_s}$, $f_D$, and
$f_{D_s}$) have to be determined precisely. In Ref.\ \cite{Chiu:2005ue},
we have presented our theoretical results of $f_D$ and $f_{D_s}$ from
lattice QCD with exact symmetry, which are in good agreement with the
recent experimental results \cite{Artuso:2005ym} from
CLEO Collaboration.

In this work, we extract the masses and the decay constants of pseudoscalar
mesons with $\s\bbar$ and $\c\bbar$, via fitting their time-correlation
functions $C_P(t)$ to the formula
\beq
\label{eq:fit_meson_asym}
F(t) = \frac{z^2}{2 m_P a}[\, e^{-m_P a t}+e^{-m_P a (T-t)}\,],
\hspace{10mm}
f_P a = (m_q + m_Q) a \frac{z}{m_P^2 a^2}
\eeq
Our results for $B_s$ are $m_{B_s} = 5385(27)(17)$ MeV,
and $f_{B_s} = 253(8)(7)$ MeV. The former is in good agreement with the PDG
value (5368 MeV). The later has not been measured in HEP experiments yet,
thus our result serves as the first prediciton from lattice QCD with exact
chiral symmetry.

In Table \ref{tab:sbbar}, we present our results of the mass spectra of the
lowest-lying states of beauty mesons with quark content $\s\bbar$. Here
we have identified the scalar $\bbar\s$ meson with the state
$B^*_{sJ}(5850)$ observed in HEP experiments, due to the proximity of their
masses. Theoretically, this implies that $B^*_{sJ}(5850)$ possesses
$J^P = 0^+$, which can be verified by HEP experiments in the future.
Further, our results of the masses of the axial-vector and tensor mesons,
which have not been observed experimentally, serves a prediction from
lattice QCD.

For the pseudoscalar meson $B_c$, we obtain
$m_{B_c} = 6278(6)(4)$ MeV, and $f_{B_c} = 489(4)(3)$ MeV. Our result of
$m_{B_c}$ is in good agreement with the experimental value 6286(5) MeV
measured by CDF Collaboration \cite{Abulencia:2005us}, but $f_{B_c}$ has
not been determined in HEP experiments. In principle, $f_{B_c}$
can be measured from the leptonic decay $B_c^+ \rightarrow l^+ v_l$, since
its decay width is proportional to $f_{B_c}^2|V_{cb}|^2$.

In Table \ref{tab:cbbar}, we summarize our results of the mass spectra
of the lowest-lying states of mesons with beauty and charm. Except for
the pseudoscalar $B_c$, other states have not been observed in experiments.
It will be interesting to see to what extent the experimental results would
agree with our theoretical values.


\begin{table}[ht]
\begin{center}
\begin{tabular}{@{}llllll@{}}
\hline
$\Qbar \Gamma q $ & $ [t_{min},t_{max}] $ & $\chi^2$/dof 
                  & Mass(MeV) & $f_P$(MeV) & PDG \\
\hline
$ \bbar\gamma_5 b $ & [27,35] & 0.72 & 9383(4)(2)   
                    & 801(7)(5)  & $ \eta_b(9300) $ \\
$ \bbar\gamma_5 c $ & [19,38] & 1.14 & 6278(6)(4)
                    & 489(4)(3)  & $ B_c(6287) $ \\
$ \bbar\gamma_5 s $ & [26,32] & 0.56 & 5385(27)(17)
                    & 253(8)(7)  & $ B_s(5368) $ \\ 
$ \cbar\gamma_5 c $ & [22,38] & 1.02 & 2980(10)(12)
                    & 438(5)(6)  & $ \eta_c(2980) $ \\ 
\hline
\end{tabular}
\end{center}
\caption{\label{tab:pmeson}
The decay constants of pseudoscalar mesons obtained in this work, together
with their masses. They are identified with the corresponding PDG mesons
listed in the last column.}
\end{table}
\vspace{-1mm}

\section{Concluding remarks}
\vspace{-1mm}

We have performed the first study of treating $\b$, $\c$ and $\s$ quarks
as Dirac fermions in lattice QCD with exact chiral symmetry. The
lowest-lying mass spectra of mesons with quark contents $\b\bbar$, $\c\cbar$,
$\s\bbar$, and $\c\bbar$ are determined, together with the pseudoscalar
decay constants (which is summarized in Table \ref{tab:pmeson}). 
Furthermore, the $ \b $ and $ \c $ quark masses are determined 
(see Table \ref{tab:mb_mc_mass}). Our results
suggest that lattice QCD with exact chiral symmetry is a viable approach
to study heavy quark physics from the first principles of QCD.

For systems with $\u$ / $\d$ quarks, one may use several quark masses in
the range $m_{u/d} < m_q < m_s$ to perform the chiral extrapolation. With
a coarser and larger lattice, say, $42^3 \times 64$ at $\beta=7.0$, it is
possible to accomodate a wider range of quark masses $m_s/4 < m_q < m_b$.
%
%
This study is now in progress.
\vspace{-1mm}


\section*{Acknowledgement}
\vspace{-1mm}

This work was supported in part by the National Science Council,
Republic of China, under the Grant No. NSC95-2112-M002-005 (T.W.C.),  
and Grant No. NSC95-2112-M001-072 (T.H.H.), and by 
the National Center for High Performance Computation, 
and the Computer Center at National Taiwan University.


\end{document}